\pdfoutput=1 

\documentclass[prb,aps,preprint]{revtex4}
\usepackage{graphicx}
\usepackage{mathrsfs}
\usepackage{dcolumn}
\usepackage{bm}
\usepackage{epsfig, amssymb}

\begin{document}
\preprint{}



\baselineskip16pt

\title{Opto-Mechanics of deformable Fabry-P\'{e}rot Cavities}


\author{Constanze Metzger}
\email{Constanze.Metzger@physik.uni-muenchen.de}
\affiliation{Center for NanoScience and Fakult\"at f\"ur Physik,
Ludwig-Maximilians-Universit\"at, Geschwister-Scholl-Platz 1,
80539 M\"unchen, Germany}

\author{Ivan Favero}
\email{Ivan.Favero@physik.uni-muenchen.de}
\affiliation{Center for NanoScience and Fakult\"at f\"ur Physik,
Ludwig-Maximilians-Universit\"at, Geschwister-Scholl-Platz 1,
80539 M\"unchen, Germany}
\author{Alexander Ortlieb}
\affiliation{Center for NanoScience and Fakult\"at f\"ur Physik,
Ludwig-Maximilians-Universit\"at, Geschwister-Scholl-Platz 1,
80539 M\"unchen, Germany}
\author{Khaled Karrai}
\email{karrai@lmu.de}
\affiliation{Center for NanoScience and
Fakult\"at f\"ur Physik, Ludwig-Maximilians-Universit\"at,
Geschwister-Scholl-Platz 1, 80539 M\"unchen, Germany}

\date{\today}

\begin{abstract}
We investigated the opto-mechanical properties of a Fabry-P\'erot
cavity with a mirror mounted on a spring. Such a structure allows
the cavity length to change elastically under the effect of light
induced forces. This opto-mechanical coupling is exploited to
control the amplitude of mechanical fluctuation of the mirror. We
present a model developed in the classical limit and discuss data
obtained in the particular case for which photo-thermal forces are
dominant.
%
\end{abstract}
%
\pacs{73.21.Hb,73.23.-b,73.23.Hk,73.20.Mf}
\keywords{}
\maketitle
%
%
\section{Introduction}
%
%
Photo-induced forces acting on a spring-mounted mirror are known
to affect its dynamics\cite{Brag77, Brag01, Brag01-2, Hoeh04,
Dor83, Marsh03, Gigan06, Arcizet06, Schliess06, Favero07, Coh99,
Mertz93, Vit02, Manc98, Kleckner06, Marquardt06, rugar07,
Vog2003}. We built a miniature Fabry-P\'erot (FP) cavity with a
moveable mirror held on a spring while the other mirror was
massive enough to be static. The flexible mirror is compliant so
that it moves under the influence of light-induced forces
originating from radiation pressure or photothermal forces that
build up in the cavity. Such forces depend on the light intensity
stored in the cavity, and their exact magnitude is determined by
the cavity's mirror separation in proportion to the optical FP
resonances. Consequently, any displacement of the mirror,
resulting for example from thermal fluctuations, leads to a change
in the light-induced force, inducing in return a change in the
mirror position. This opto-mechanical coupling is referred to as
intrinsic light-induced back-action\cite{Brag77}.
\\
An optical back-action mechanism shifting the resonance frequency
and adding damping on a mechanical resonator was first reported by
V.B. Braginsky\cite{Brag77} three decades ago. Optical back-action
remained a field of interest, especially in the research area of
gravitational wave detection\cite{Brag01, Brag01-2}. Gravitational
wave detectors, mostly Michelson interferometers (for example
LIGO, a Michelson interferometer with arm lengths of 4 km that is
illuminated with a 6 W Nd:YAG laser beam\cite{LIGO}), are prone to
get unstable because of optical back-action. Instabilities were
reported as well in smaller scale systems. A centimeter sized
mirror hung on strings and serving as one mirror of a FP cavity
showed mechanical instability under few Watts of
illumination\cite{Dor83}. More recently, back-action was reported
in microscale
systems\cite{Vog2003, Schliess06}\\
When the photon back-action force is delayed in time with respect
to changes in mirror position, additional dissipation in the
mirror's motion occurs without adding any additional mechanical
fluctuations. The enhanced dissipation leads to reduced
vibrational fluctuation and temperature of the mirror\cite{Hoeh04,
Gigan06, Arcizet06, Schliess06, Favero07}, a situation referred to
as passive optical cooling\cite{Hoeh04}.
Quantum mechanical behavior of a miniature mirror is
expected\cite{Marsh03} when the optical cooling becomes efficient
enough to cool the mirror near its vibrational ground state.
Experiments using a combination of photo-thermal forces and
radiation pressure to cool a micromirror passively reach a
temperature range of about 10 K in references\cite{Hoeh04,
Gigan06, Arcizet06}.
Optical cooling dominated by radiation pressure has been
demonstrated not only in FP cavities\cite{Gigan06, Arcizet06} but
in silica microtoroids\cite{Schliess06} with a diameter in the
range of 100 $\mu$m as well. Unfortunately, optical cooling
mechanisms start to become inefficient as soon as the mirror
reaches size smaller than the diffraction limit of light in the
cavity. Nevertheless, cooling of a micromirror with a diameter in
the range of the laser wavelength
was recently successfully demonstrated\cite{Favero07}.\\
In analogy to optical cooling, capacitative cooling of a
nano-mechanical resonator through charge coupling with a
superconducting single-electron-transistor was shown\cite{Naik06}.
For a reviewpaper see ref\cite{search06}.\\
In a pioneering work and in contrast to passive cooling
mechanisms, Cohadon, Heidmann and Pinard demonstrated the
possibility of optical active cooling using an external electronic
feedback loop in their system\cite{Coh99}. In an earlier set of
data by Mertz and coworkers\cite{Mertz93}, optical induced damping
by active feedback was observed. In cold damping schemes, a laser
beam is directed towards the flexible mirror and can displace it
exerting radiation pressure\cite{Coh99} or a photo thermal
force\cite{Mertz93}. The velocity of the mirror is detected and
the laser intensity is adjusted by an electronic feedback loop in
an appropriate way\cite{Vit02, Manc98}. In principle, because this
technique modulates the light intensity in proportion to a signal
derived from the mirror amplitude noise, it adds technical
fluctuations in the system. Using active optical cooling, up to
now effective temperatures as low as 135 mK could be
reached\cite{Kleckner06} with a cantilever starting from room
temperature. Recently, active cooling of a cantilever from 2.2 K
down to about 3 mK was
observed\cite{rugar07} using not optical but electrostatic feedback forces.\\
In this paper, we present a model describing passive optical
cavity cooling in a classical approximation and report on the
passive cavity cooling of a micromirror by photo-thermal
back-action forces under various experimental conditions.
\\
\\
In chapter \ref{sec:1}, we present solutions to the equation of
motion of a mirror with a delayed light-induced force acting on
it. A derivation of the vibrational temperature of a mirror cooled
by photo-induced forces is given in chapter \ref{sec:2}. Chapter
\ref{sec:3} describes the mirror's equation of motion under a
weakly modulated light-induced force. Different micro FP
experiments giving rise to optical cooling are presented in
chapter \ref{sec:4} and \ref{sec:5}. Finally, in chapter
\ref{sec:6} we compare the cooling power for different light
induced forces. We discuss the possibility that cooling by
photo-thermal effects allows reaching lower temperatures compared
to cooling by radiation pressure.
%
%
\section{Equation of motion under constant illumination}\label{sec:1}
\begin{figure}[htb]
  \includegraphics[width=10cm, clip]{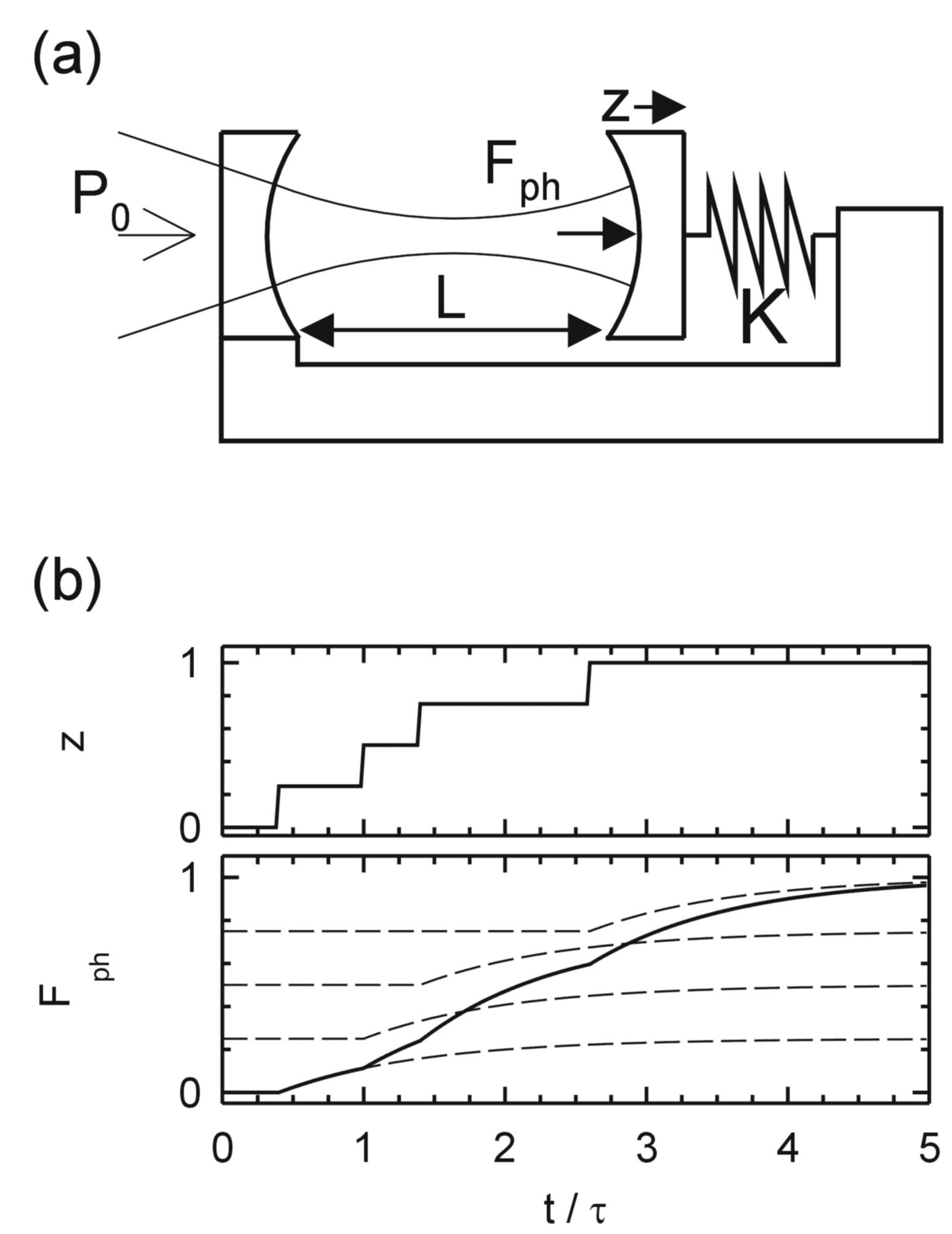}
  \caption[fig:modell-cavity]{(a) schematic model of a deformable Fabry-P\'erot cavity.
  (b) After discreet step-shaped changes in mirror distance z,
  the light-induced force F  grows after a characteristic delay time $\tau$.}
   \label{fig:modell-cavity} 
\end{figure}

%
In this chapter, we solve the equation of motion of a vibrating
harmonic oscillator forming a mirror of a FP cavity in the limit
of small vibrational amplitudes.
In our setup, a laser beam is coupled into the cavity through a
fixed semi-transparent input mirror. Depending on the mirror
distance, a resonance builds up in the cavity. The photons stored
in the deformable FP cavity exert a force $F_{ph}$ on the
compliant mirror originating on the light-field present in the
cavity. The force can be any photon-induced force such as
radiation pressure, photo-thermal deformation of the mirror,
radiometric pressure or else. For sake of generality $F_{ph}$ in
our analysis is assumed to be any possible photon induced force
that is proportional to the local light intensity at the location
of the mirror. Generally such forces do not respond
instantaneously at a change in mirror position, but only delayed
after a characteristic time constant $\tau$. For example, the
finite photon storage time of a cavity accounts for the delay of
radiation pressure forces with respect to a change in cavity
lenght, while photo-thermal action on the mirror is retarded by
the time it takes to conduct heat conduction along the cantilever.
A model system with a mirror that is able to move under the
influence of a delayed photon force is shown in
FIG.\ref{fig:modell-cavity} (a).
\\
We consider the equation of motion for the center of mass position
$z$ of a oscillator with an effective mass $m$, mechanical damping
$\Gamma$ and spring constant $K$. The mirror thermal fluctuations
are assumed to be driven by an thermal Langevin Force $F_{th}$.
\begin{equation}\label{eq:oscillator}
m\ddot z(t)+m\Gamma \dot z(t) +K z(t)=
F_{th}(t)+F_{ph}\left(z(t)\right)\,.
\end{equation}
In the following, we model the total light induced force on the
cantilever. To illustrate,we consider that the cantilever position
fluctuates in random increments under the effect of thermal
excitations. The photon force responds retarded in time. After a
step of $z_n-z_{n-1}$ at time $t_n$, the light-induced force
$F_{ph}$ follows with the delay time $\tau$ as depicted in FIG.
\ref{fig:modell-cavity} (b). If we were to stop the random motion
of the mirror at step $n$, the light-induced force would reach
asymptotically the static value $F(z_n)$. To model the behavior of
$F_{ph}(z(t))$ after $N$ steps in mirror position we sum up all
force increments such
\begin{equation}
\label{eq:force} F_{ph}(z_{N}(t))=
 F(z_0)+\sum_{n=1}^{N}h(t-t_n)[F(z_n)-F(z_{n-1})]
\end{equation}
where the function $h(t)$ describes the time delay. This discrete
sum can be reformulated as a continuous integral in time
\begin{equation}
\label{eq:force2} F_{ph}(z(t))= F(z_0)+\int_{0}^t dt'\,\frac{d
F(z(t'))} {d t'}h(t-t').
\end{equation}
The equation of motion  we need to solve then reads as
\begin{equation}\label{eq:eqofmotion-before-LT}
m\ddot z(t)+m\Gamma \dot z(t) +K z(t)= F_{th}(t)+F(z_0)+\int_{0}^t
dt'\,\frac{d F(z(t'))}{d t'}h(t-t').
\end{equation}
This equation\cite{Hoeh04} leads to complex dynamics with
multi-stability points treated in a recent work by F. Marquardt
and coworkers\cite{Marquardt06}. Here we focus on optical cooling,
so for all practical purpose we assume the mirror amplitudes to be
small compared to the change in cavity length needed for the
optical resonance condition to change substantially. In terms of
the FP cavity finesse $\emph{F}=(\pi /2)\, g$ with
$g=2\sqrt{R}/(1-R)$ this constraint translates into
$z<<\lambda /(2\pi g)$ where $R$ is the reflectivity of the cavity mirrors.\\
Equation (\ref{eq:eqofmotion-before-LT}) is solved by Laplace
transform, which is defined for a function $f(t)$ as
\begin{equation}\label{eq:def_of_LT}
%
                            f_\omega  =  \int_{0}^{\infty}dt\,f(t)\, e^{-i\omega
                            t}\,.
\end{equation}
The constant force term $F(z_0)$ in eq.
(\ref{eq:eqofmotion-before-LT}) has no time dependence and simply
leads to a static shift of the oscillator's average position. By
selecting the new average position for $z$ it can be dropped from
eq. (\ref{eq:eqofmotion-before-LT}). The Laplace transform of eq.
(\ref{eq:eqofmotion-before-LT}) yields
\begin{equation}\label{eq:LT}
%
 -m\omega^2 z_\omega+i \omega m \Gamma z_\omega+Kz_\omega= \int_{0}^{\infty}dt\, e^{-i\omega
t}\left[F_{th}(t)+\int_{0}^{t}dt'\,\frac{dF(z(t'))}{dt'}h(t-t')\right]
\end{equation}
As $F(z(t'))$ depends on time indirectly through $z(t')$, its
derivative in eq.(\ref{eq:LT}) is rewritten as
\begin{equation}dF(z(t'))/dt'=
\frac{\partial F(z(t'))}{\partial z} \frac{\partial
z(t')}{\partial t'}.
\end{equation}
In accordance with the small amplitude approximation, $F(z(t'))$
is developed in a Taylor expansion around $z(t_0)$:
$F(z(t'))\approx F(z(t_0))+[z(t')-z(t_0)]\nabla F$ where we used
the abbreviation $\partial F(z(t'))/\partial z|_{z=z(t_0)}=\nabla
F$. In the small amplitude fluctuation approximation, the partial
derivative $\partial F(z(t'))/\partial z$ is now approximated with
$\nabla F$. We can reformulate eq. (\ref{eq:LT}) as follows
\begin{equation}\label{eq:LT-1}
 -m\omega^2 z_\omega+i \omega m \Gamma z_\omega+Kz_\omega=F_{th, \omega}+
  \int_{0}^{\infty}dt\, e^{-i\omega
t}\left[\int_{0}^{t}dt'\nabla F\frac{\partial z(t')}{\partial
t'}h(t-t')\right]\,.
\end{equation}
With the property of Laplace transform for convolutions
\begin{equation}
\int_0^\infty dt\, e^{-i\omega t}\left[ \int_0^t dt'
f_1(t')f_2(t-t') \right]=f_{1, \omega}f_{2, \omega}
\end{equation}
eq. (\ref{eq:LT-1}) is reformulated as
\begin{equation}\label{eq:eqmx}
 -m\omega^2 z_\omega+i \omega m \Gamma z_\omega+Kz_\omega= F_{th, \omega}
 + \nabla F i \omega z_\omega h_\omega\,.
\end{equation}
We assume that the shape of the delay function is of exponential
type
\begin{equation}
h(t)=1-e^{-t/\tau}\,.
\end{equation}
This is reasonable, because $h(t)$ describes the timescale the
cavity system needs to approach a new equilibrium state after a
disturbance. For instance radiation pressure reacts with an
exponential behavior. The other process considered in this work,
the heat flow in an absorbing mirror after a change of cavity
length, has an exponential response as well. The Laplace transform
of the response function $h(t)$ is given by
\begin{equation}h_\omega=\frac{1}{i \omega (1+i \omega \tau)}\,.
\end{equation}
%
%
%
The terms on the right hand side of eq. (\ref{eq:eqmx}) can be
regrouped in powers of $\omega$ and eq. (\ref{eq:eqmx}) is
rewritten as
\begin{equation}-m\omega^2 z_\omega+i \omega m \Gamma_{\mathrm{eff}} z_\omega+K_{\mathrm{eff}}z_\omega=F_{th,
\omega}
\end{equation}
with an effective damping
\begin{equation}\label{eq:gammaeff}
\Gamma_{\mathrm{eff}}= \Gamma\left(1+Q_M\frac{\omega_0 \tau}
{1+\omega^2 \tau^2}\frac{\nabla F}{K}\right)
\end{equation}
and an effective spring constant
\begin{equation}\label{eq:keff}
K_{\mathrm{eff}}= K\left(1-\frac{1}{1+\omega^2 \tau^2}\frac{\nabla
F}{K}\right).
\end{equation}
In eq. (\ref{eq:gammaeff}), we used the vibrational harmonic
resonance frequency of the center of mass of the mirror
$\omega_0^2=K/m$ and we defined the mechanical quality factor such
that
\begin{equation}\label{eq:def_Q}Q_M=\frac{\omega_0}{\Gamma}\,.
\end{equation}
Both the effective damping and rigidity are unusual in that they
now include a frequency dependent term. The frequency dependency
is that of a low-pass filter that ensures that at very high
frequencies the retarded back-action has no effect on the
properties of the harmonic oscillator. Above cut-off the
oscillating mirror behaves as if it was placed in the dark. In the
limit of low frequencies (static limit) the effective damping and
spring rigidities are constant and as a result the solution of the
equation of motion is that of an harmonic oscillator with
optically modified frequencies and quality factor. For
applications involving laser cooling of the lowest mechanical
vibrational mode, the frequency range of interest is
$\omega\approx\omega_0$, the cantilever's resonance frequency. We
define the effective resonance frequency
\begin{equation}\label{eq:feff}
\omega_\mathrm{eff}^2=\omega_0^2\left(1-\frac{1}{1+\omega^2
\tau^2}\frac{\nabla F}{K}\right)\,.
\end{equation}
where $\omega_\mathrm{eff}^2=K_\mathrm{eff}/m$.
The solution for the amplitude in the frequency domain of the
harmonic oscillator is
\begin{equation}\label{eq:solution-light}z_\omega=\frac{F_{th,
\omega}}{m}\frac{1}{\omega_{\mathrm{eff}}^2-\omega^2+i\omega\Gamma_{\mathrm{eff}}}.
\end{equation}
It is important to note that we did not take into account that
$h(t)$ is a function of the cavity detuning in contrast to the
model in ref\cite{Arcizet06}. In our simplified approach with low
finesse cavities the effect of detuning on $h(t)$ is not
measurable but becomes significant at high
finesses\cite{Schliess06, Arcizet06}. The delay time of
photo-thermal forces is entirely determined by heat conduction in
the mirror and is not dependent on cavity detuning at all.
%
\section{Effective Temperature}\label{sec:2}
\begin{figure}[h!]
 \centering
  \includegraphics[width=15cm, clip]{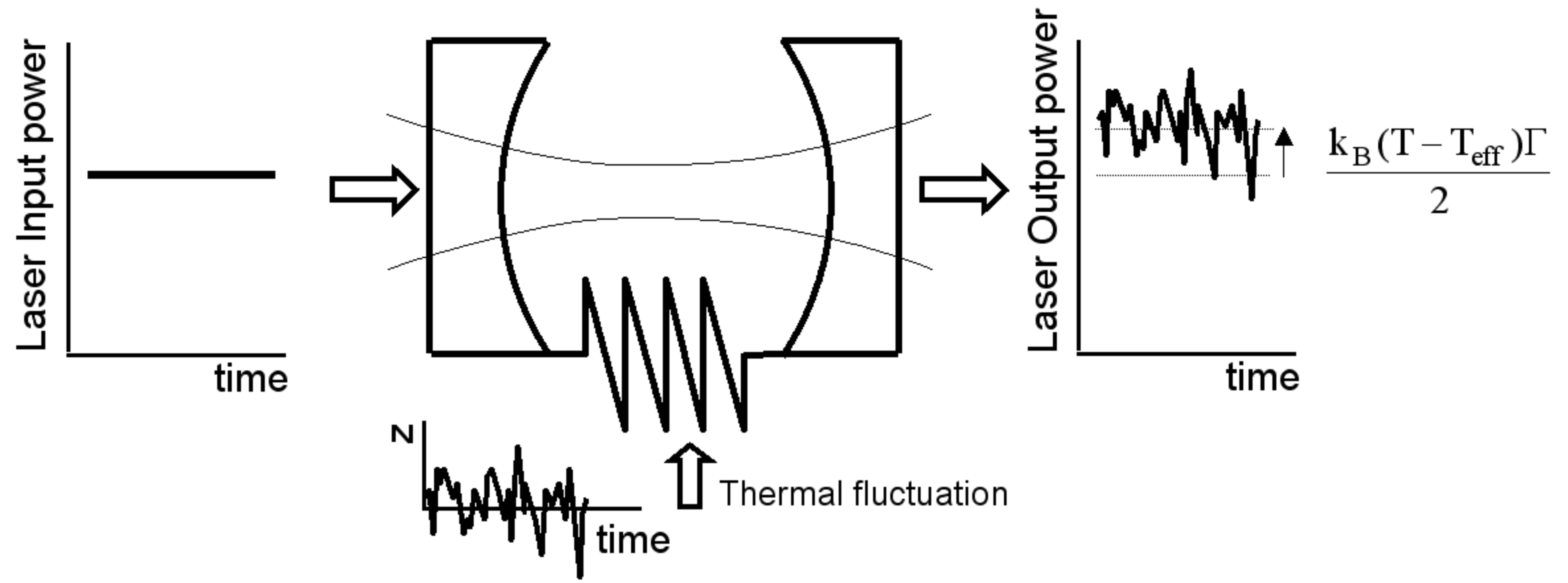}
  \caption[fig:energyloss]{A FP cavity with a mirror attached on a
  spring is illuminated with a laser beam. The
  input laser intensity is assumed to be noiseless. The
  transmitted light shows amplitude fluctuations,
  that are impressed on the original amplitude by the thermal
  fluctuation of the mirror. More importantly, the transmitted laser light has
  an averaged intensityenhanced by the fluctuations added by the mechanical resonator.
  The mirror vibrational motion has been
  cooled and the excess energy turns into photons.}
   \label{fig:energyloss} 
\end{figure}

%
%
In thermodynamical equilibrium without illumination and any
light-induced effects, the average power in the mechanical ground
mode of the mirror center of mass motion is described by the
equipartition theorem:
\begin{equation}\label{eq:equipartition}\frac{1}{2}K\int_0^\infty dt\,|z_{\mathrm{dark}}(t)|^2=\frac{1}{2}k_BT\,.
\end{equation}
Here $k_B$ is Boltzmann's constant and $T$  the bath temperature.
An important property of Laplace transforms is that the integrated
Laplace coefficients $\int d\omega|z_\omega|^2$ equals the time
average
\begin{equation}\label{eq:timeaverage_is_omegaaverage}
\int_0^\infty d\omega|z_{\mathrm{dark}, \omega}|^2 =
\int_0^\infty    dt\,|z_{\mathrm{dark}}|^2\,.
\end{equation}
This expression provides the prescription for performing
vibrational thermometry, namely a method to extract a temperature
from the measurement of the spectral distribution of the Brownian
motion of the mirror. First the rigidity $K$ must be determined
independently, for instance by measuring the resonance frequency
knowing the oscillator effective mass, then the spectrum of the
fluctuation amplitude $z_\omega$ is measured on a sufficiently
extended frequency range around the vibrational resonance
frequency and averaged over a large enough number of measurements.
Finally the integration of $|z_\omega|$ multiplied by the rigidity
gives the thermal energy experienced by the harmonic oscillator
and hence the temperature. We will use this prescription later on
to determine the temperature of the mirror coupled to the optical
cavity. The expression for the frequency averaged square modulus
of the amplitude can be now computed using the solution $z_\omega$
of eq. (\ref{eq:solution-light}) but still as a function of the
still non-explicitly expressed thermal fluctuation force component
$F_{th, \omega}$. In absence of light in the cavity the
equipartition theorem gives us already the opportunity to derive
the expression of $F_{th, \omega}$ that we can then finally use to
obtain the dynamics of the mirror with light in the cavity. As we
will see shortly, the result will be that the mirror fluctuates in
a way nearly identical to the Brownian motion of the original
harmonic oscillator in dark but with a modified temperature
induced by the presence of light in the cavity. In dark, setting
all light induced effects to zero in eq. (\ref{eq:oscillator}) for
$z_\omega$, we have
\begin{equation}\label{eq:oscillator-dark}z_{\mathrm{dark},
\omega}=\frac{F_{th, \omega}}{m}\frac{1}{\omega_0^2-\omega^2+i
\omega \Gamma}\,.
\end{equation}
With the reasonable assumption that the spectral force density of
thermal vibrations given by $F_{th,\omega}$ are equally
distributed over all frequencies, one can calculate the strength
of the thermal force. We assume that
\begin{equation}\label{eq:driving-fluct}|F_{th, \omega}|^2=S df
\end{equation}
in every frequency interval $df$ with a constant spectral density
$S$ which can be calculated in the next step by integrating eq.
(\ref{eq:oscillator-dark}) over all frequencies $\omega$
\begin{equation}\label{eq:oscillator-dark1}\int_0^\infty d\omega|z_{\mathrm{dark},\omega}|^2=
 \int_0^\infty d\omega\frac{S}{2\pi m^2}\frac{1}{(\omega_0^2-\omega^2)^2+\omega^2\Gamma^2}\,.
\end{equation}
The experimentally relevant assumption $\Gamma<<\omega_0$ is made,
so the integral simplifies to
\begin{equation}\label{eq:oscillator-dark1-1}\int_0^\infty d\omega|z_{\mathrm{dark},\omega}|^2
=\frac{S}{2\pi m^2 \Gamma^2 \omega_0^2} \int_0^\infty
d\omega\frac{1}{4(\frac{\omega_0-\omega}{\Gamma})^2+1}\,.
\end{equation}
leading to the solution 
\begin{equation}\label{eq:zquared-dark}\int_0^\infty d\omega|z_{\mathrm{dark}, \omega}|^2 =
 \frac{S}{4 K \Gamma m}.
\end{equation}
With that result, the solution of the oscillator's spectrum eq.
(\ref{eq:zquared-dark}) can be inserted in the equipartition
theorem eq. (\ref{eq:equipartition}). The driving fluctuation eq.
(\ref{eq:driving-fluct}) is determined:
\begin{equation}\label{eq:langevin-T}|F_{th, \omega}|^2=4k_B
T m\,\Gamma\frac{d\omega}{2\pi}\,.
\end{equation}
Finally the thermal noise spectrum of a harmonic oscillator in the
dark is
\begin{equation}\label{eq:brownian-noise}|z_{\mathrm{dark}, \omega}|^2=\frac{4 k_B T \Gamma}{m}
\frac{1}{(\omega_0^2-\omega^2)^2+(\omega
\Gamma)^2}\frac{d\omega}{2\pi}\,.
\end{equation}
Now, we still need to find an expression for the thermal driving
force $F_{\mathrm{th}, \omega}$ in the solution of the equation of
motion with light eq. (\ref{eq:solution-light}). When the light is
turned on, the spectral force density $F_{th, \omega}=Sdf$ is not
influenced by the photon induced force and eq.
(\ref{eq:langevin-T}) still holds, because it is only dependent on
the natural mechanical damping $\Gamma$ and the undisturbed spring
constant $K$. The spectral amplitude of a mirror under
illumination is
\begin{equation}\label{eq:solution-light-squared}|z_\omega|^2=
\frac{4k_BT \Gamma}{m}
\frac{1}{(\omega_{\mathrm{eff}}^2-\omega^2)^2+(\omega\Gamma_{\mathrm{eff}})^2}\frac{d\omega}{2\pi}\,.
\end{equation}
Integrating this over all frequencies and using the property of
Laplace transforms eq. (\ref{eq:timeaverage_is_omegaaverage})
gives
\begin{equation}\label{eq:zquared}\int_0^\infty dt\,|z|^2 =
\frac{\Gamma}{\Gamma_{\mathrm{eff}}} \frac{k_B
T}{K_{\mathrm{eff}}}.
\end{equation}
This averaged squared amplitude is related to a temperature
$T_{\mathrm{eff}}$ via the equipartition theorem:
\begin{equation}\label{eq:equipartition_eff}\frac{1}{2}K_{\mathrm{eff}}\int_0^\infty dt\,
|z|^2=\frac{1}{2}k_BT_{\mathrm{eff}}\,.
\end{equation}
Solving this for the effective temperature and using eq.
(\ref{eq:zquared}) yields
\begin{equation}\label{eq:teff}
\frac{T_\mathrm{eff}}{T}=\frac{\Gamma}{\Gamma_{\mathrm{eff}}}\,\,.
\end{equation}
No absorption of light in the mirror was taken into account up to
now. Still even dielectric mirrors possess a residual absorption
leading to heating. If the temperature is increased considerably
above the bath temperature, eq. (\ref{eq:teff}) needs to be
corrected. The bath temperature $T$ has to be substituted then
with the temperature the mirror would reach in absence of optical
cooling $T+\Delta
T$.\\
In a previous work\cite{Hoeh04}, we established that
$T_\mathrm{eff}/T=(\Gamma/\Gamma_{\mathrm{eff}})( K/
K_\mathrm{eff})$ which does not take into account that the
effective temperature is determined by the squared noise amplitude
$\int_0^\infty dt\,|z|^2$ multiplied with the independently
measured effective spring constant $K_\mathrm{eff}$ instead of the
unperturbed spring constant $K$.
This correction creates a factor $K_\mathrm{eff}/K$ yielding the effective temperature eq. (\ref{eq:teff}).\\
With the help of eq. (\ref{eq:gammaeff}), the result of eq.
(\ref{eq:teff}) is reformulated as
\begin{equation}\label{eq:temp}
\frac{T_{\mathrm{eff}}}{T}=\frac{1}{1+Q_M\frac{\omega_0
\tau}{1+\omega^2\tau^2}\frac{\nabla F}{K}}
\end{equation}
revealing the physical parameters playing a role in cavity cooling.\\
The cooling stops when the static spring constant
$K_{\mathrm{eff}}(\omega=0)$ reaches zero and becomes negative. At
this point, mirror bistability sets in\cite{Vog2003} and no stable
measurement is possible any more. Consequently, a theoretical
limit of cooling is obtained for $K_{\mathrm{eff}}=K(1-\nabla
F/K)=0$ in eq. (\ref{eq:temp}) and considering the optimal case of
$\omega_0 \tau=1$
\begin{equation}
\frac{T_{\mathrm{eff, Limit}}}{T}=\frac{1}{1+Q_M/2}\,.
\end{equation}
This expression shows that the mechanical quality factor $Q_M$,
which relates to the ability of the mechanical mode to dissipate
its energy, plays a central role for the
optical cooling mechanism.\\
According to eq. (\ref{eq:teff}) the lowest effective temperature
is entirely driven by the damping modified through the cavity
effect. In turn this modification in damping exists only if a time
delay exists between the motion of the mirror and the resulting
change in the light induced force it experiences, see eq.
(\ref{eq:gammaeff}). So the essence of optical cooling finds its
root on the retarded back-action on the
mirror displacement.\\
Up to now, we did not offer an explanation as to where the thermal
energy extracted from the vibrating cantilever goes. It turns into
fluctuation of the electromagnetic field escaping the cavity as
shown in FIG. \ref{fig:energyloss}.
The system formed by the mechanical oscillator and the
electromagnetic field remains at constant temperature. We offer a
possible picture on how this happens. The fluctuating cavity
length modulates the photon frequency at all frequencies but with
amplitude maxima at the vibrational resonance frequency. Such
amplitude modulation of the light-field produces side bands above
and below the photon frequency with peaks shifted on both sides by
the vibrational resonance of the mirror (in Raman spectroscopy
they would be Stokes and anti-Stokes resonances). When the laser
is red detuned from the cavity transmission maximum, the band with
shorter wavelength is closer to the transmission peak. Seen from
the outside world a detector would measure a fluctuating
irradiance imbalance between the side bands as more blue shifted
light is reaching the detector than red shifted. This excess of
energy is given by the difference in transmitted light power
between the blue and red side of the band and this over the
typical delay time constant for the light induced force to correct
against the mirror fluctuation. The excess energy has been taken
away from the very source that produced the side bands to begin
with, namely from the Brownian fluctuation of the mirror. In this
picture, the cooling is optimal when the frequency width of the
cavity, that is the inverse storage time $1/\tau$, is comparable
to the side-band frequency separation from the laser light
frequency, in other words when $\omega_0 \tau\approx1$. This
picture seems to be consistent with the model and in particular it
is easy to see that with a zero time delay the net excess energy
is also zero and no cooling is possible. An alternative picture
possibly more appropriate to photo-thermal cooling is the
following. The laser light is tuned to be red-shifted from a
transmission peak of the cavity. When the cavity length fluctuates
and say becomes shorter over a certain time period, the
transmission peak gets closer to the laser line and more light can
be stored in the cavity during that time. The result of the excess
light is to exert more pressure on the mirror as to oppose the
cavity from becoming even shorter. In the opposite case, when the
cavity gets longer upon a thermal fluctuation, the averaged steady
state light pressure that displaced the mirror from its position
in dark reduces and the mirror tends to move back under its own
restoring elastic force as to oppose this very fluctuation. The
retarded back-action makes this force oppose the mirror velocity
$dz/dt$ and not only its instantaneous position $z$. It is
therefore a dissipative force and during the typical response
time, energy is irreversibly lost to the light field outside the
cavity. In this picture the cavity serves as a reservoir of energy
stored in form of light, and the rate of energy leakage from this
reservoir is fully controlled by the mirror kinetics. Energy
conservation dictates that mechanical energy can be transformed
into energy that escapes the reservoir in form of light.\\
%
%
%
\section{Equation of motion under modulated illumination}\label{sec:3}
%
The solution of the equation of motion of a mirror under the
influence of light-induced forces of chapter \ref{sec:1} is
generalized for a weakly modulated light-induced force. This
modification proves to be useful, because a measurement with
modulated laser light opens up the possibility to measure the
magnitude of the light-induced force as well as its delay time.
The technique makes is possible to determine if either radiation
pressure, photo-thermal pressure or even a summation of both
effects are responsible for the observed cooling effects. We took
advantage of this method in a modulated laser measurement that
is discussed in chapter \ref{sec:4}.\\
If the laser intensity is weakly modulated, the light-induced
force is described by
\begin{equation}\label{eq:modulation}F(z(t),t)=(1+\varepsilon(t))F_{ph}(z(t))
\end{equation}
with a small modulation strength $\varepsilon(t)<<1$. The
light-induced force has now an explicit dependence on $t$ and
differs from eq. (\ref{eq:force2}) as follows
\begin{equation}
\label{eq:forceext}
 F(z(t),t)=  F_{ph}(z_0)+\int_{0}^{t}\left(\frac{\partial F_{ph}}{\partial t'}+\frac{\partial F_{ph}}{\partial z}
 \frac{\partial z}{\partial t'}\right)h(t-t')dt'\,.
\end{equation}
The solution for the amplitude 
is:
\begin{equation}\label{eq:solution-drive}z_\omega=\left(\frac{F_{th, \omega}}
{m}+\frac{F_{ph}}{m}\frac{\varepsilon_{\omega}}{1+i\omega
\tau}\right)\frac{1}{\omega_{\mathrm{eff}}^2-\omega^2+i\omega\Gamma_{\mathrm{eff}}}\,.
\end{equation}
Compared to the solution without external excitation of the mirror
eq. (\ref{eq:solution-light}), the amplitude has an additional
term $(F_{ph}/m)\,\varepsilon_\omega /(1+i\omega \tau)$. This term
offers a way to extract both the delay time $\tau$ and the
magnitude of the light-induced force $F_{ph}$ from a measurement
of the real part as well as the imaginary component of the
response $z_\omega$ so the measurement can be done with the aid of
lock-in detection (see chapter \ref{sec:4}) by measuring the in
and out of phase component of the reflected light. Using eq.
(\ref{eq:solution-drive}) to model the data, the delay time of the
force is extracted. Besides, if different light-induced forces
like radiation pressure and photo-thermal pressure are present in
the setup, the ratio of different forces can be determined when
their response times differ significantly.
\\
In the next two chapters, we are investigating in two different
setups for the optical cooling of the vibration modes of a gold
coated AFM silicon cantilever. In this system, the presence of the
bilayer gives rise to a photo-thermal bending of the lever under
illumination. The delay time of the light-induced force is the
time of thermal response of the lever.
%
\section{Cooling of the ground mode}\label{sec:4}
%
\begin{figure}[h!]
 \centering
  \includegraphics[width=15cm, clip]{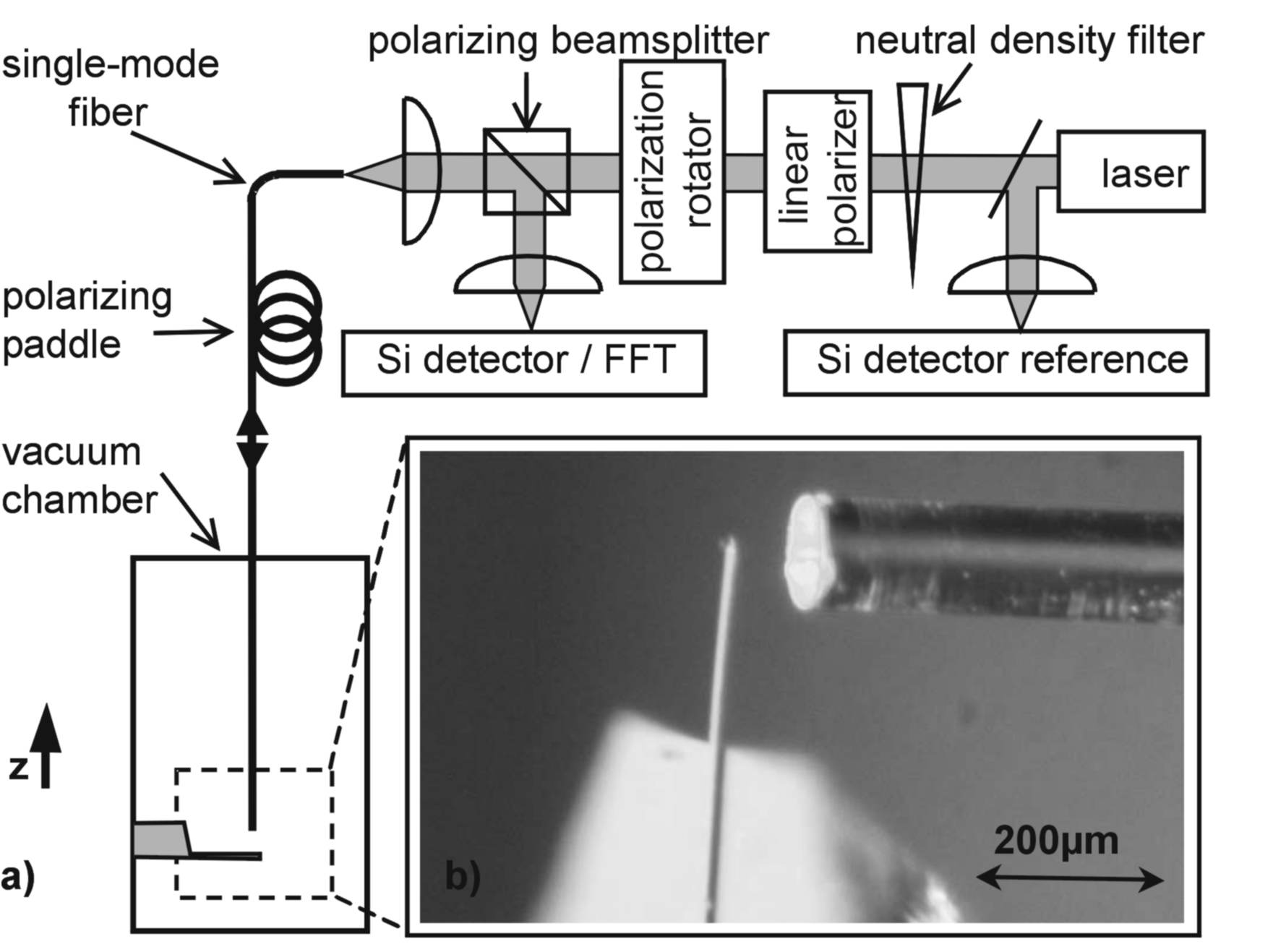}
  \caption[fig:photo-cavity]{(a) Schematic of the experimental setup. Inset (b) shows a photograph
  of a similar cavity as used in the experiment but with a lever to mirror separation
  greatly increased for the picture. During measurements, the cavity length was 34 $\mu$m.}
   \label{fig:photo-cavity} 
\end{figure}
In this chapter, a setup displaying passive back-action cooling is
shown (see FIG. (\ref{fig:photo-cavity})). We used alternatively a
red HeNe-laser (Research electro optics LHRP 1701, $\lambda=$632.8
nm, 17 mW) or a diode laser ($\lambda=$670 nm, 5 mW) beam coupled
into a single mode optical fiber (numerical aperture 0.13). The
highly coherent HeNe laser was used for the vibrational resonance
linewidth measurements shown in FIG. \ref{fig:fig2}. For
measurements involving laser amplitude modulation, we preferred
using the diode laser because it could be easily modulated. A
neutral density filter wheel allowed tuning the laser power
continuously over almost four orders of magnitude. The reflected
laser power was measured at the level of the Si detector was
varied from 35 nW to up to 150 $\mu$W. The fiber was introduced
into a vacuum chamber operating down to a pressure in the
$10^{-6}$ mbar range. Reaching this low enough pressure was
important in order to reduce the damping of the cantilever as
shown later in FIG. \ref{fig:fig2} (a). The fiber end forming a
cavity mirror in the vacuum chamber was thoroughly polished and
coated with a gold film of 19 nm by thermal evaporation under high
vacuum. A silicon cantilever (Nanosensors) with a width of 22
$\mu$m, a thickness of 0.47 $\mu$m, a length of 220 $\mu$m and a
spring constant of 0.008 N/m was mounted at a distance of 34
$\mu$m of the polished fiber end. Gold layers of 36 nm were
deposited on each side of the lever. A simulation of the coated
cantilever optical properties gave a reflectivity of 82$\%$ for a
laser wavelength of 633 nm. The distance between fiber and
cantilever was tuned by applying a DC-voltage between them to
create a capacitative force. About 15 V were required in order to
detune the cavity through three resonances. The light reflected
from the cavity was coupled back into the fiber. A fiber paddle
polarizator was used to rotate the linear polarization of the
reflected light in order to be directed by a polarizing
beamsplitter onto a Si-photodetector and minimize back reflected
light on the laser. We increased this way the collected efficiency
by a factor four. For additional isolation we used a polarization
rotator ($\lambda/2\pm 1 \%$ Fresnel rhombus, B. Halle Nachf.,
400-700 nm) rotating the linear laser polarization by $45^{o}$ per
pass and a linear
polarizer (Glan-Thompson, isolation 50 dB) before fiber coupling.\\
\begin{figure}[h!]
 \centering
  \includegraphics[width=13cm, clip]{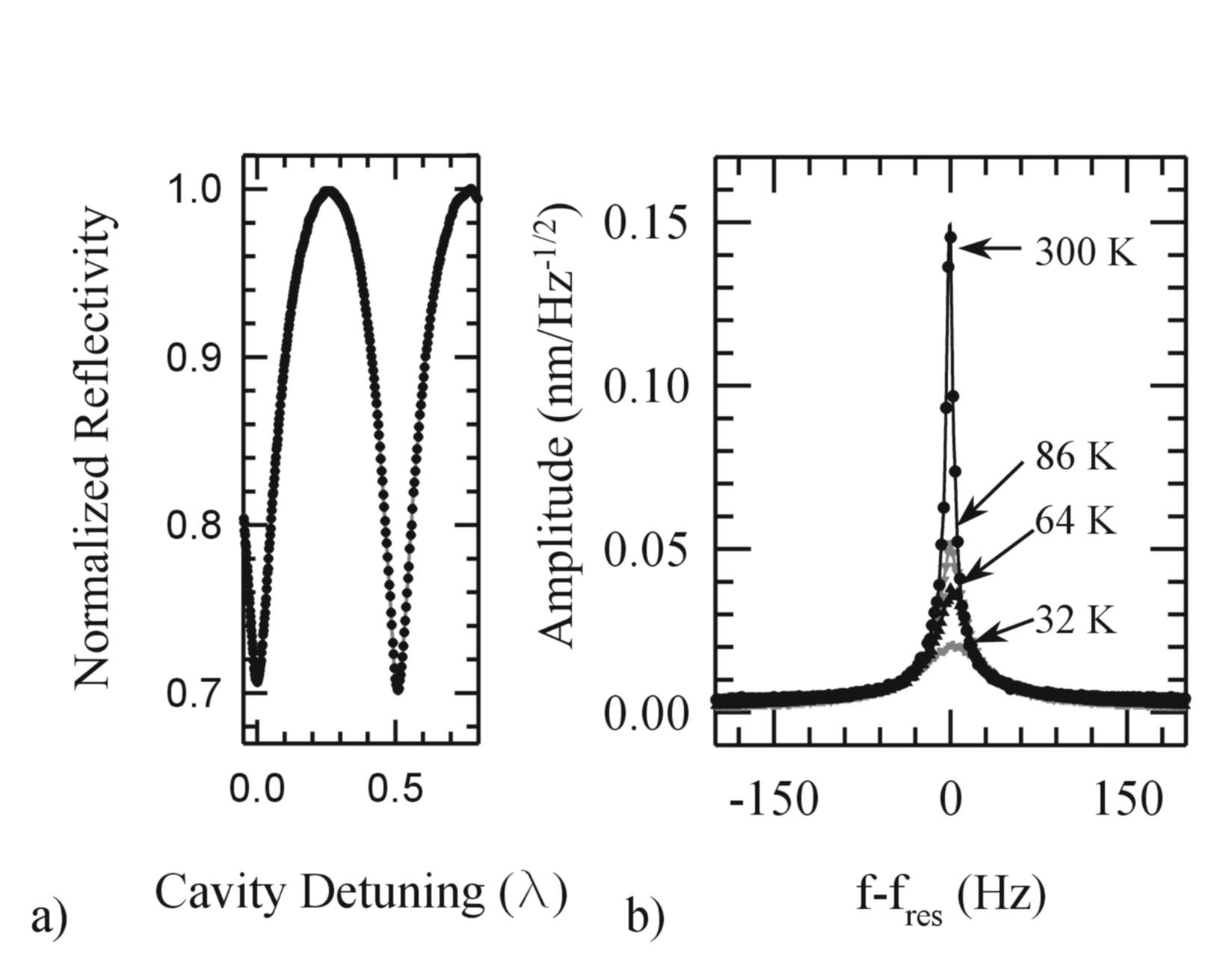}
  \caption[fig:coolingcant]{(a) Normalized reflectivity of plan-plan
  cavity setup shown in FIG. \ref{fig:photo-cavity}. The cavity detuning
  is calibrated in units of the wavelength
  $\lambda$. The Finesse is 4,
  with the parameter g=2.5.
  (b) Amplitude fluctuation spectral density near vibrational resonance of the
  cantilever with $f_0=7.3$ kHz with different laser powers taken at cavity detuning of $+\lambda/25$
  from a cavity resonance. The largest amplitude corresponds to thermal
  fluctuation at 300 K with mechanical damping $\Gamma=28$ Hz, measured with
  reflected laser power of 3.1$\mu$W. The other measurements correspond to reflected laser
  powers of 0.87 mW, 1.3 mW, and 3.6 mW with damping of $\Gamma=$ 98 Hz,
  131 Hz, and 263 Hz. The effective temperatures of the spectra
  are from top to bottom 300 K, 86 K, 64 K, and finally 32 K.}
   \label{fig:coolingcant} 
\end{figure}
In FIG. \ref{fig:coolingcant} (a), the normalized reflectivity of
the FP cavity is shown. The cavity finesse is about four, which
corresponds to $g=2.5$. FIG. \ref{fig:coolingcant} (b) shows
spectra of the cantilever fundamental harmonic at 7.3 kHz with
effective temperatures of 300 K, 86 K, 64 K, 32 K respectively.
All curves are taken at the same cavity detuning of $\Delta
z=+\lambda /(2\pi g \sqrt{3})\approx\lambda /25$ for which one
expects maximum gradient of the light-induced force, and therefore
maximum cavity cooling\cite{Vog2003}. At very low reflected laser
power of 3.1 $\mu$W one measures the amplitude fluctuation
spectral density near vibrational resonance of the cantilever
corresponding to a temperature of 300 K. A fit with eq.
(\ref{eq:brownian-noise}) is obtained with the parameters $f=7265$
Hz, $K=2.5\times10^{-2}$ N/m, $\Gamma=28$ Hz. At increased laser
power of 3.6 mW the effective damping is found to be
$\Gamma_\mathrm{eff}=263$ which relates using eq. (\ref{eq:teff})
to an effective temperature of 32 K for this set of data. So far
the lowest temperature obtained with this setup\cite{Hoeh04} was
18 K. In order to achieve highest possible cooling effect,
different parameters
have to be optimized as stated in eq. (\ref{eq:temp}). \\
First, the mechanical damping $\Gamma$ of the cantilever needs to
be minimized. The damping of a resonator includes several
contributions such as clamping losses, defects in crystal
structure, surface losses\cite{Mih95} and damping due to
scattering of air molecules to name a few. The latter can be
reduced by running the system in vacuum. A simple model of the gas
damping can be found by assuming that the viscous damping by
molecular scattering is $F_\mathrm{visc}=(N m_\mathrm{N}
v)/t_{scat}$ with $N$ the number of atoms scattering off the
cantilever, $m_\mathrm{N}=4.6\times 10^{-26}$ kg the mass of
nitrogen atoms, $v=510$ m/s the mean atomic velocity at 300 K and
$t_{scat}$ the mean scattering time. This approximation predicts
that at a pressure of $10^{-3}$ mbar already molecular scattering
should account for 1$\%$ of the damping. In reality, we still see
a sizeable change in quality factor going from $10^{-3}$ mbar
($\Gamma=61.7$ Hz) to $5\times10^{-6}$ mbar ($\Gamma=14.5$ Hz).
The linewidth at full width half maximum FWHM relates to the
mechanical damping such as $\mathrm{FWHM}=\sqrt{3}\Gamma/(2 \pi)$.
We find a linewidth of 17 Hz corresponding to a quality factor of
$Q_M=744$ for $10^{-3}$ mbar and a linewidth of 4 Hz ($Q_M=3161$)
for $5\times10^{-6}$ mbar as shown in FIG. \ref{fig:fig2} (a) at
low reflected power. Evidently, the observed damping cannot be
explained by molecular viscous damping alone. Molecular adsorption
on the cantilever surface may be responsible for the additional
damping so at lower pressure desorption could explain the improved
quality factor. As seen in eq. (\ref{eq:gammaeff}), the linewidth
of the mechanical resonance is modified linearly with laser power
as long as the photon-induced force is linear with intensity. In
the cooling regime, it is broadened with increasing laser power
starting from the natural linewidth at dark. In FIG.
\ref{fig:fig2} (a) the linear dependency of the linewidth with the
reflected laser power is plotted in logarithmic scale for
different chamber pressures, showing
smallest possible linewidth at low pressure and low laser power.\\
%
%
%
%
%
%
In order to maximize the cooling efficiency, a tradeoff between
the reflectivity of the cantilever and its mechanical damping had
to be made. Higher reflectivity should increase the cavity finesse
and therefore lead to stronger cooling effect, through both an
increase of he light power circulating in the cavity close to
resonance and an increase of its gradient upon position.
Unfortunately, increasing the reflectivity by evaporating a
thicker gold layer on the cantilever adds additional mechanical
damping as well\cite{Sandberg05}. In our experiment we used
different thicknesses of evaporated gold on many cantilevers of
the same kind, the quality factor decreased by an
order of magnitude as shown in FIG. \ref{fig:fig2} (b).\\
\begin{figure}[h!]
 \centering
  \includegraphics[width=15cm, clip]{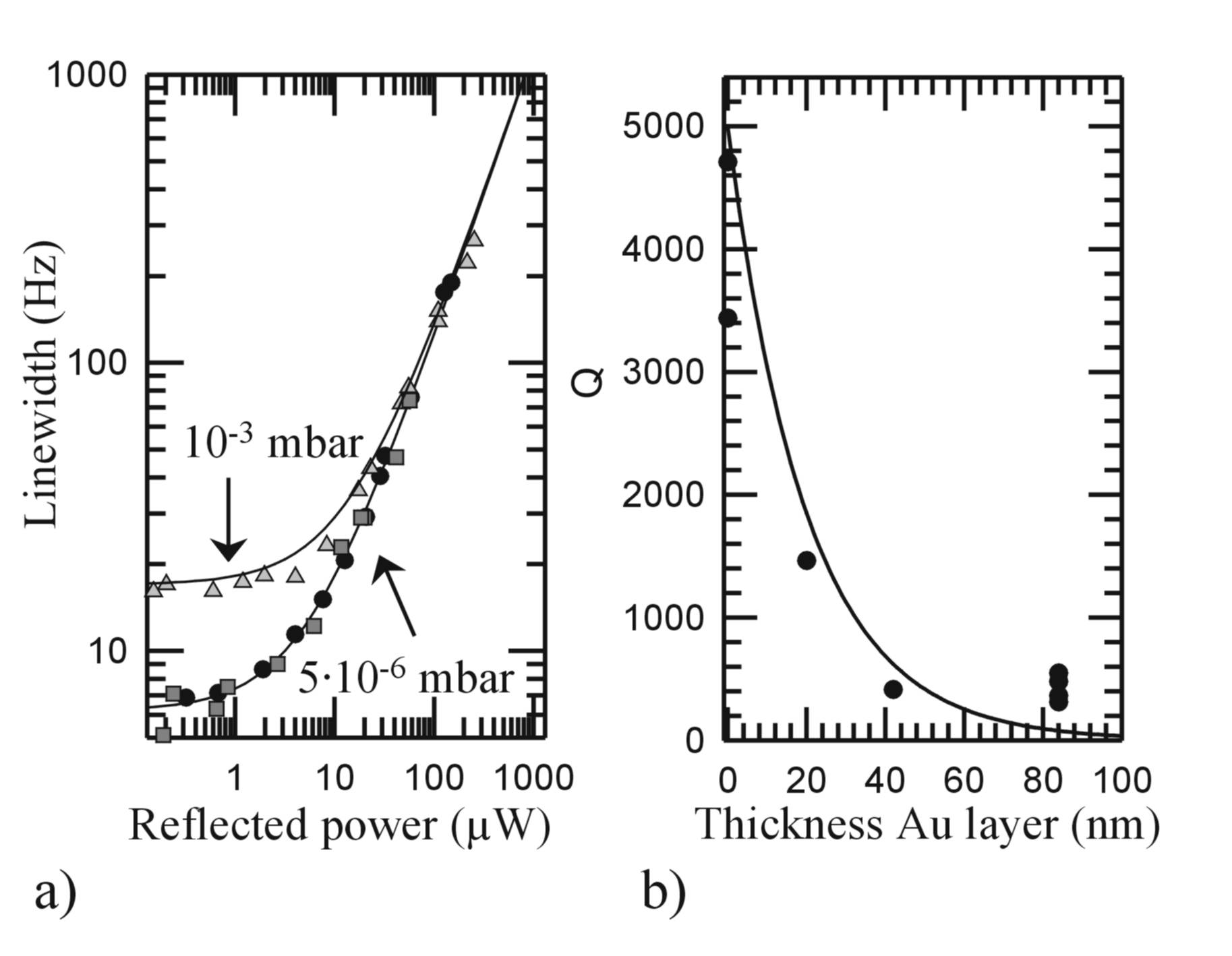}
  \caption[fig:fig2]{(a) Dependence of the cantilever's
  resonance linewidth at full width half maximum with pressure and laser power.
  The triangles show data taken with a HeNe-laser
  at moderate pressure of about $10^{-3}$ mbar, while the circles
  where taken at minimum pressure of $5\times 10^{-6}$ mbar.
  The squares were taken also at minimum pressure but using
a red (670 nm) diode laser instead. Clearly,
  the linewidth of the cantilever is much smaller at
  $5 \times 10^{-6}$ mbar. (b) Mechanical
  quality factors of different cantilevers with various
  thicknesses of evaporated gold. The grey line is a guide to the eye.}
   \label{fig:fig2} 
\end{figure}
Second, to enhance the cooling efficiency the parameter $\tau$
needs to be optimized. An inspection of eq. (\ref{eq:temp}) shows
that optimum cooling is reached for $\omega \tau =1$. In case of
thermal bending of the cantilever, the delay time of the
light-induced force is given by the time it takes the thermal
energy to diffuse along the cantilever. For a bilayer cantilever
consisting of a thin gold layer with thickness $u_{Au}$ and a
silicon layer of thickness $u_{Si}$, this thermal diffusion time
constant $\tau_{pth}$ can be approximated by\cite{Bar1994,
Gim1994}
\begin{equation}\label{eq:reaktionszeit}
\tau_{pth}=l^2 \,\frac{\rho_{Si} c_{Si} u_{Si}+\rho_{Au} c_{Au}
u_{Au}}
 {\Lambda_{Si} u_{Si}+\Lambda_{Au} u_{Au}}.
\end{equation}
with $\rho$ the density $c$ the specific heat capacity, $\Lambda$
the thermal conductivity and $l$ the length of the cantilever.
Taking the parameters of the cantilever given above and
$\rho_{Si}=2.33$ g/cm$^3$, $\rho_{Au}=19.3$ g/cm$^3$,
$c_{Si}=0.71$ J/(gK), $c_{Au}=0.128$ J/(gK), $\Lambda_{Si}=1.48$
W/(cm K) and $\Lambda_{Au}=3.17$ W/(cm K) one finds
$\tau_{pth}=0.5$ ms. With the mechanical resonance frequency of
the cantilever of $f_0=7.3$ kHz, a value of $\omega_0 \tau=25$ is
found. It is interesting to note that $\omega_0 \tau$ is a
function of material thickness alone. The resonance frequency of a
multi layer cantilever is given by
\begin{equation}\label{eq:multi}
\omega_0=\frac{(1.875)^2}{l^2}\sqrt{\frac{1}{u_1\rho_1+u_2\rho_2}\int_{-u/2}^
{u/2}E(u-u_0)^2 du}
\end{equation}
where $u_0$ denotes the cantilever's neutral stress axis. The
Young modulus $E$ is integrated over the thickness $u$ of the
different cantilever layers\cite{Sand05}. For a cantilever
consisting of one layer, eq. (\ref{eq:multi}) simplifies to
$\omega_0=u/l^2\sqrt{E/\rho}$ and the corresponding thermal
constant is $\tau_{pth}=l^2/h$ where $h=\Lambda/(\rho c)$ is the
thermal diffusivity. Setting the condition $\omega \tau = 1$ leads
to an optimal thickness $u_\mathrm{opt} = h \sqrt{\rho/E}$.
For silicon at room temperature, this optimal thickness is found
to be 10 nm with $h=8.6 \times 10^{-5}$ m$^2$/s, the values were
found in\cite{PropSi}. This value is far too small for fabrication
of free standing silicon structures. However, a diamond resonator
with optimized thickness seams feasible. With
$E_\mathrm{diam}=1.1\times10^{12}$ N/m$^2$,
$\rho_\mathrm{diam}=3200$ kg/m$^3$ and
$h_\mathrm{diam}=5.09\times10^{-4}$ m$^2$/s, one finds an optimal
thickness of 27.5 nm. A resonator with that thickness and a length
of 900 nm would feature a resonance frequency of 100 MHz. For a
silicon cantilever, the temperature can be used to tune $\omega_0
\tau$ since the specific heat $c$ and the thermal conductivity
$\Lambda$ show strong temperature dependence and a temperature
where
 $\omega_0 \tau=1$ can be found. For example, the
 diffusivity of silicon\cite{PropSi} increases by a factor
 of 20 from 300 K to 80 K, so placing the cantilever
 of our experiment at liquid nitrogen temperature of 77 K
 should allow reaching the optimal condition $\omega_0 \tau\approx 1$ in contrast
 to $\omega_0 \tau\approx 25$ at room temperature.\\
For radiation pressure induced cooling, the delay time is given by
the cavity storage time for a photon\cite{Hoeh04}
$\tau_R=L/(c(1-R))$. With our parameters we find that the cavity
storage time is in the range of 0.2 ps and therefore orders of
magnitudes smaller than the thermal diffusion time constant. For
this reason in this experiment we expect optical cooling to be
mostly dominated by photo-thermal effects and not by radiation
pressure.
\begin{figure}[h!]
 \centering
  \includegraphics[width=15 cm , clip]{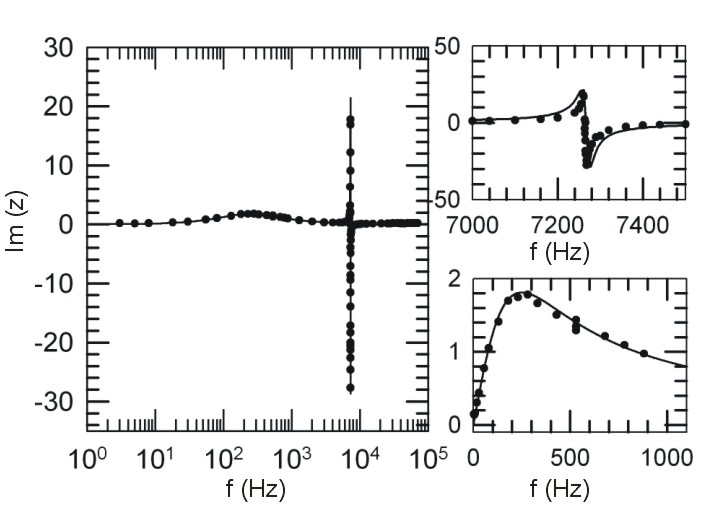}
  \caption[fig:fig3]{Response measurement of driven cantilever amplitude.
  Frequency sweep from 0 to 100 kHz, the small picture above shows a zoom of the region around the
  cantilever resonance frequency at 7.3 kHz .
  Inset at the bottom shows a zoom of the enhanced response at f=284Hz arising at the frequency
  where $\omega \tau =1$. The response frequency corresponds to $\tau_{pth}=560\mu$s.}
   \label{fig:fig3} 
\end{figure}
In order to obtain the measured value $\tau_{pth}$ we performed a
response measurement of the cantilever's motion driven by a weakly
modulated laser light-induced force.
The laser intensity of a red diode laser with a wavelength of 670
nm was modulated weakly by modulating the laser current with a
signal generator and we used the internal reference of a lock-in
(SR 7265). About 5$\%$ of the overall intensity was modulated such
that the modulation parameter $\varepsilon$ in
eq.(\ref{eq:modulation}) was 0.05. The modulation frequency was
swept in single steps in the frequency range from DC to 100 kHz.
The reflected signal measured at the Si-photodetector was
demodulated using the lock-in. We were interested in measuring the
imaginary part of the overall amplitude response shown in eq.
(\ref{eq:solution-drive}).
The measurement of the real part of eq. (\ref{eq:solution-drive})
for low laser amplitude is
\begin{equation}\label{eq:real-part}
\mathit{Re}(z_\omega)=\frac{\varepsilon_\omega
F_\mathrm{pth}}{m}\frac{\omega_0^2-\omega^2(1+\Gamma\tau)}
{[(\omega_0^2-\omega^2)^2+\omega^2\Gamma^2](1+\omega^2\tau^2)}
\end{equation}
without taking into account the contribution of $F_{\mathrm{th}}$
which is much smaller than $F_\mathrm{pth}$. The real part is
always superimposed with the amplitude of the modulated light
intensity $\varepsilon P R$. This adds a complication in detecting
the direct opto-mechanical effect. In contrast, the measurement of
the imaginary part
\begin{equation}\label{eq:im-part}
\mathit{Im}(z_\omega)=\frac{\varepsilon_\omega
F_{pth}}{m}\frac{-\omega[(\omega_0^2-\omega^2)\tau+\Gamma]}
{[(\omega_0^2-\omega^2)^2+\omega^2\Gamma^2](1+\omega^2\tau^2)}
\end{equation}
is purely dependent on the opto-mechanical
response\cite{Favero07}. In the experiment, we found two different
competing forces. A photo-thermal force with a delay time in the
range of heat diffusion time was coexistent with the quasi
instantaneous radiation pressure force.\\
The imaginary part shows a characteristic local maximum where the
response function $\omega\tau/(1+\omega^2\tau^2)$ is maximal at
the frequency $1/(2 \pi \tau)$. In a modulated response
measurement, one measures an overall phase shift occurring in the
system. The phase shift is not only caused by the cantilever's
response alone but also includes the phase shifts in the detection
apparatus. To solve this technical problem, we devised a measuring
procedure cancelling spurious phase shift effects at all
frequencies. For each measurement at a given modulation frequency,
we first measured the spurious phase shifts by switching off all
signal coming from the opto-mechanical response of the cantilever
itself. This is obtained when the force gradient $\nabla F=0$, so
we tuned the cavity such that the reflectivity was maximum. The
phase is then set to zero at the lock-in. In a next step, without
changing any other parameter, we detuned the cavity to a regime of
maximum $\nabla F$. At this point, the imaginary component of the
signal is solely originating from the cantilever opto-mechanical
response. For each modulation frequency we repeated the procedure
explained above. The result is shown in FIG.\ref{fig:fig3}. We
were able to fit the data with eq. (\ref{eq:im-part}) using a
combination of two forces acting on the lever. The first is a
thermal bending force with a time delay of $\tau_{pth}=560\,\mu$s.
The second is the quasi-instantaneous ($\tau \approx 0$) radiation
pressure that does not contribute here to cooling. The ratio of
the forces was found to be $F_{pth}/F_{rad}=-95$. On resonance,
$\nabla F_{pth}/(1+\omega_0^2\tau^2)$ is the contribution of the
thermal force to the light-induced frequency shift. Its magnitude
is found to be 95/625=0.15 smaller than the contribution of
$F_{rad}$ so effects on frequency shift in this experiment were
dominated by radiation pressure alone\cite{Hoeh04}.
\\
The modulated experiment shown in FIG. \ref{fig:fig3} demonstrated
convincingly that the observed cooling effects were dominated not
by radiation pressure but by a photo-thermal bending force that
was 95 times stronger than radiation pressure and had an opposite
sign. The value found experimentally for the delay time $\tau=560
\,\mu$s is in agreement with the prediction of 0.5 ms made with
the help of eq. (\ref{eq:reaktionszeit}). This indicates that a
small asymmetry in the thickness of the gold layers on the two
faces of the cantilever creates a thermal force opposing the
radiation pressure. The imaginary response shows a clear maximum
at the cantilever resonance and an enhancement at the the
frequency $f=1/(2\pi \tau)=284$ Hz corresponding to the thermal
response of the system. We see that the cooling effect at the
cantilever's ground mode of 7300 Hz is not optimal, because
$\omega_0 \tau \approx 25$ is far from one. As mentioned earlier,
placing the lever at 77 K should optimize the cooling to $\omega_0
\tau$ about 1.
%
\section{Simultaneous cooling of the fundamental vibrational mode and its first harmonic}\label{sec:5}
%
In an experiment using a cantilever with a gold coating on one
side only, much stronger thermal forces were measured. Here, we
used a slightly different cavity arrangement designed to increase
the cavity finesse as well as to decrease the size of the
laser beam on the microlever.\\
\begin{figure}[h!]
 \centering
  \includegraphics[width=13cm, clip]{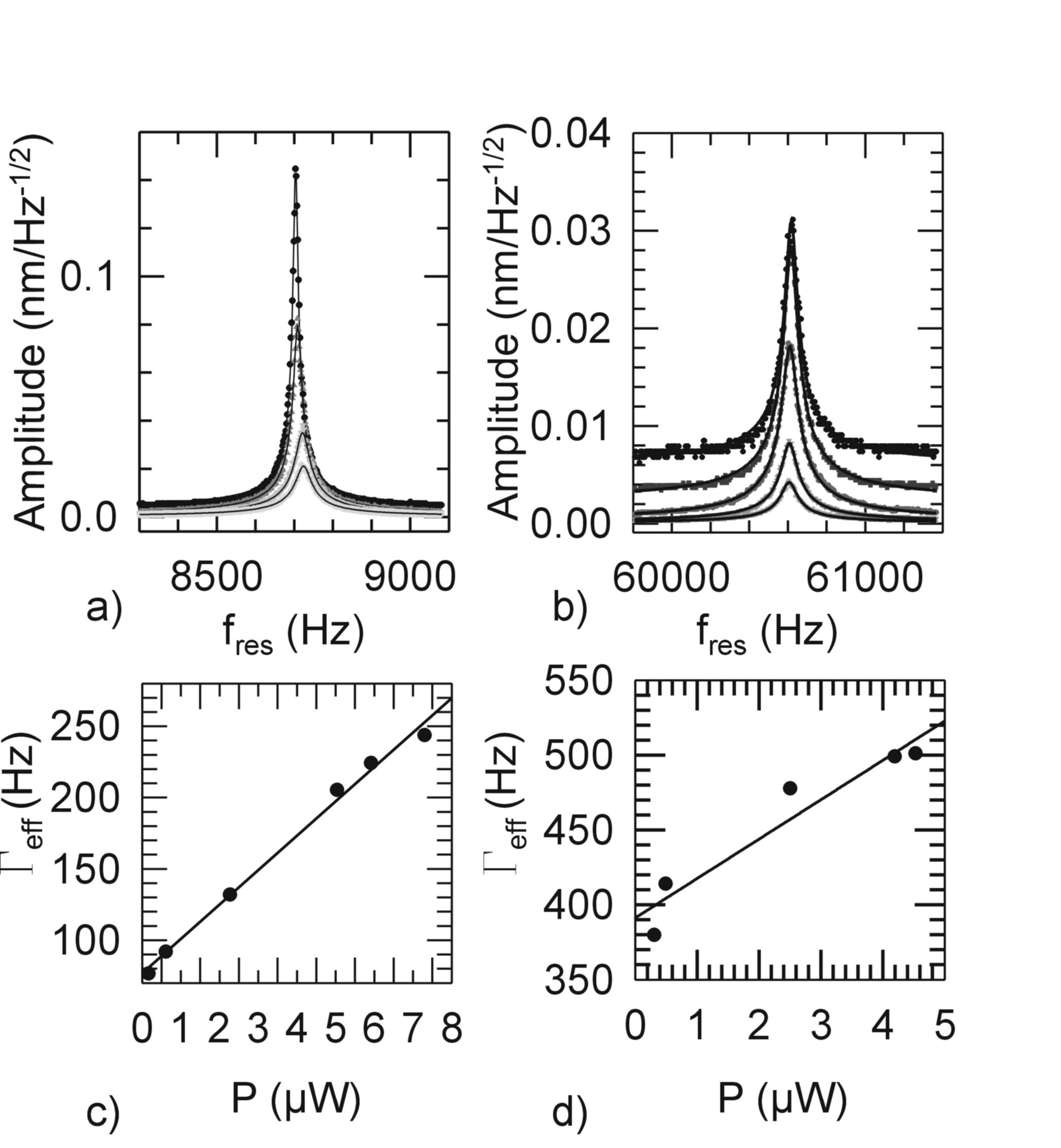}
  \caption[fig:fig4]{(a) Cooling behavior of fundamental
  vibrational mode at 8.7 kHz. Laser powers coupled in the fiber before the cavity are 0.16$ \mu$W for
  Brownian peak, then 2.25$ \mu$W, 5.8$ \mu$W, 7.6$ \mu$W, corresponding to 300 K, 174 K, 102 K, and
  94 K respectively. The fits were made according to eq. (\ref{eq:solution-light-squared}).
  The effective damping for the spectra is
  shown in (c).
  (b) cooling of first harmonic at 60.6 kHz, laser powers 0.31 $ \mu$W
0.49 $ \mu$W, 3.14 $ \mu$W, 4.19 $ \mu$W, 4.53 $\mu$W
corresponding to 300 K, 290 K, 251 K, 240 K, 239 K. The offset of
the spectra shows $1/\sqrt{P}$ dependence and is caused by shot
noise of the laser.
(c) Effective damping $\Gamma_\mathrm{eff}$ with laser power
before cavity for the ground mode at 8.7 kHz.
$\Gamma_\mathrm{eff}$ shows linear power dependence according to
eq. (\ref{eq:gammaeff}).
d) Effective damping $\Gamma_\mathrm{eff}$ with laser power before
cavity for the first harmonics at 60.6 kHz. }
   \label{fig:fig4} 
\end{figure}
The light of a red monomode HeNe-laser (Sios, $\lambda=$633 nm,
1.3mW) was coupled into a single mode fiber (NA=0.13). The fiber
end was polished and coated with a reflecting gold layer of 30 nm
(yielding a reflectivity of 70$\%$) to form the first cavity
mirror. The divergent beam coming out of the fiber was collimated
with a first lens with numerical aperture of NA=0.25 (Geltech
glass aspheric lens, diameter 7.2 mm, focal length 11.0 mm), then
refocused on the sample with a second lens identical to the first
one. The microscope yielded a gaussian focus on the sample with a
$1/e^2$ diameter of 6 $\mu$m. This diameter includes 86$\%$ of the
gaussian light mode. The sample is a cantilever with length 223
$\mu$m, thickness 470 nm, width 22 $\mu$m, spring constant
$K=0.01$ N/m and a gold layer of 42 nm this time on one side only.
A simulation of the silicon-gold bilayer system gave a
reflectivity of 91$\%$.  The cavity finesse defined by the sample
and the fiber end was $F=8$. FIG. \ref{fig:fig4} (a) shows cooling
of the cantilever's first mode of vibration at 8.7 kHz from room
temperature down to 94 K. The lowest effective temperature of 94 K
was reached with the laser intensity of 7.6$\mu$W (power coupled
into fiber before first cavity mirror). This is by far not the
maximal achievable power with the used laser. However, the cooling
was limited by the
appearance of instabilities in the static spring constant\cite{Vog2003}.\\
A response measurement with weakly modulated laser done with the
same procedure as described in chapter \ref{sec:4} gave a value
for the thermal diffusion time of $\tau=760\,\mu$s and a ratio of
$F_{pth}/F_{rad}\approx 4000$.
An interesting point concerning photo-thermal cooling is shown in
FIG. \ref{fig:fig4} (b). The figure shows photo-thermal induced
cooling of the cantilever's first harmonic, measured under the
same conditions as the cooling of the ground mode shown in FIG.
\ref{fig:fig4} (a). This simultaneous cooling of two modes is very
much consistent with the fact that the energy lost to the lowest
vibration mode does not feed another mechanical mode of the
cantilever but is transferred out of the system.
%
%
\section{Photo-thermal versus radiation pressure cooling}\label{sec:6}
%
In this chapter, we compare the lowest temperature reached with
photo-thermal cooling and radiation pressure cooling. Both cooling
methods are considered in optimal cooling condition at $\omega_0
\tau =1$. At present time it is not obvious which method will lead
to the lowest temperatures in the quest for quantum ground state
cooling of a mechanical resonator. Photo-thermal cooling on one
side is always accompanied with optical absorption in the
resonator limiting the system's temperature. Its advantages
nevertheless are apparent, because the light-induced force can be
orders of magnitudes stronger than radiation pressure and the
condition $\omega_0 \tau=1$ can be reached by careful design of
the gold layer on the cantilever or else by adjusting the bath
temperature as shown in chapter \ref{sec:4}. In radiation pressure
cooling on the other hand, the system still experiences residual
absorption heating up the resonator. Additionally, radiation
pressure is by far not as strong as photo-thermal forces. To
obtain a strong radiation pressure force, the light intensity in
the cavity has to be increased considerably leading
in turn to increased absorption heat input to the resonator. \\
First, we address the situation of ideal cooling with radiation
pressure without any residual absorptions. As derived in chapter
\ref{sec:2}, the effective temperature is given by eq.
(\ref{eq:temp}). In order to reach the minimum effective
temperature, the cavity is tuned to the maximal gradient of
radiation pressure\cite{Vog2003}
\begin{equation}
\nabla F_{rad, max}\approx \frac{2 P_0}{c\lambda}2\sqrt{R}g^2
\end{equation}
where $P_0$ is the laser power sent on the cavity. This maximal
light-induced force gradient occurs at a detuning of $\lambda /(2
\pi g \sqrt 3)$ from a cavity resonance\cite{Vog2003}. With this
expression and using the cavity storage time
$\tau_{rad}=L/(c(1-R))$ the minimal effective temperature is
found:
\begin{equation}\label{eq:tmin}
\frac{T_\mathrm{eff, rad}}{T}\approx\left( 1+
\frac{1}{1+\omega_0^2 \tau_{rad}^2} \frac{P_0}{mc^2 \Gamma}
g^3\frac{L}{\lambda /2} \right)^{-1}.
\end{equation}
Here, $L$ is the cavity length. For a setup with $\omega_0
\tau_{rad}=1$ this simplifies to
\begin{equation}\label{eq:teff-rp}
\frac{T_{\mathrm{eff}, rad}}{T}=\left(\frac{P_0 }{2mc^2
\Gamma}\frac{L}{\lambda/2}g^3 \right)^{-1}
\end{equation}
for strong cooling $T_{\mathrm{eff}}<<T$. No absorption of light
in the mirror was taken into account up to now. Lowest
temperatures can be achieved by increasing laser power and
finesse, or else by choosing a system with low mass and
damping as well as a large cavity length.\\
Now, we analyze cooling by to photo-thermal forces.
This effect is not only due to differential thermal expansion in a
multilayered composite mirror surface, but can also originate from
a non-uniform temperature distribution around the region where
light is absorbed. In both cases the effect is not instantaneous
and leads to time constants usually much larger than a single pass
time of flight of photons through the cavity. The effect can be
seen as an effective force that displaces the mirror in proportion
to the amount of absorbed laser power. In order to compare
photo-thermal forces with radiation pressure we introduce an
effective index $n$ that accounts for a photo-thermal induced
force $F_{pth}$ that would scale like $nF_{rad}$, where
$F_{rad}=2P_0R/c$ is the force resulting from radiation pressure
acting on the mirror. Since the photo-thermal force relies on the
absorption $\alpha$ in the mirror, $n$ is proportional to
$\alpha$. For illustration, the factor $n$ for the doubly-sided
gold coated cantilever is -95, while for the cantilever coated on
one side only it is 4000.
\\
For better analogy with radiation pressure, where the delay time
$\tau_{rad}$ scales as $L/(c(1-R))$, we give the photo-thermal
retardation time in units of $\tau_{rad}$ such that
$\tau_{pth}=n_{\tau}\tau_{rad}$. Physically, $n_{\tau}$ represents
the thermalization time constant of the mirror in units of
$\tau_{rad}$. For the first experiment shown in chapter
\ref{sec:4}, this parameter is $2.8\times10^{9}$, for the second
in chapter \ref{sec:5},
it is around $1.9\times10^{6}$.\\
With these definitions, the minimal effective temperature for
photo-thermal cooling can be formulated with the help of eq.
(\ref{eq:temp}) in the approximation of $\omega_0 \tau=1$ and for
strong cooling $T_{\mathrm{eff}}<<T$:
\begin{equation}\label{eq:teff-pth}
\frac{T_{\mathrm{eff, pth}}}{T}=\left(n n_\tau \frac{P_0
}{2mc^2\Gamma}\frac{L}{\lambda/2}g^3\right)^{-1}.
\end{equation}
We stress that the effective indexes $n$ and $n_\tau$ are purely
phenomenological. They are introduced here to allow a direct
comparison between photo-thermal and radiation pressure cooling in
terms of the ultimate cooling temperatures they yield.\\
We are now able to compare directly the minimal reachable
temperatures and cooling power $P_{cool}$
accounted for by radiation pressure and photo-thermal forces.\\
In dark and at thermal equilibrium the lever's mechanical
fluctuation dissipates its energy $k_BT/2$ at a rate $\Gamma$. The
dissipated power is therefore $(k_BT/2)\Gamma$ and is in
equilibrium with the power that feeds the fluctuation as dictated
by the fluctuation-dissipation theorem. When the mechanical
resonator is cavity cooled, its vibrational effective temperature
is $T_\mathrm{eff}$ but at the same time the internal source of
mechanical dissipation $\Gamma$ is still present. In other words
the internal mechanical dissipation rate that heats the resonator
is still $\Gamma$. When the vibrational mode reaches a
steady-state at a temperature $T_\mathrm{eff}$, the heat load in
the mirror is $(k_BT_\mathrm{eff}/2)\Gamma$. Consequently, in
order to maintain a steady state end temperature, the optical
cooling extracts energy from the fluctuations in the mirror at a
rate
\begin{equation}
P_\mathrm{cool}=\frac{k_B (T-T_\mathrm{eff})}{2}\Gamma.
\end{equation}
Making use of eq. (\ref{eq:teff}), we obtain
\begin{equation}
P_\mathrm{cool}=\frac{k_BT}{2}\Gamma\left(1-\frac{\Gamma}{\Gamma_\mathrm{eff}}
\right).
\end{equation}
For large temperature differences that are typical for efficient
cooling we have $T_\mathrm{eff}<<T$ which translates into
$\Gamma_\mathrm{eff}>>\Gamma$. The maximum cooling power is then
approximated by $(k_BT/2)\Gamma$, which is interestingly still
thermal mechanical fluctuation of a resonator in the dark.\\
Until now, we did not consider any absorptions in the mechanical
resonator. Yet, real mirrors always have a finite absorption that
acts as a heat source and leads to added fluctuation of the
vibrational mode. As a result, it limits the lowest achievable
temperature. The absorbed light heats the mirror body to reach a
new temperature $T+\Delta T$ where the excess temperature $\Delta
T = \beta(\alpha P_\mathrm{mirror})$ is proportional to $\alpha
P_\mathrm{mirror}$, the amount of absorbed laser power at the
location of the mirror. Here, $\beta$ is a proportionality factor
that translates the absorbed power to an excess temperature and is
dependent on the mirror's heat conduction and geometry properties.
In a FP cavity, the laser power at the location of the mirror is
larger than the laser power outside the cavity by an amount
$P_\mathrm{mirror}= gP_0$ proportional to the cavity finesse. The
excess temperature accounted for by residual absorption in the
mirror corresponds to a heating power $P_\mathrm{heat}=(k_B\Delta
T/2)\Gamma$ of the vibrational mode that ultimately balances the
cooling power. The maximum laser power $P_{max}$ usable before the
absorption counteracts the cooling is obtained by equating
$P_\mathrm{cool} = P_\mathrm{heat}$ which gives
\begin{equation}
P_{max}=\frac{T}{\alpha \beta
g}(1-\frac{\Gamma}{\Gamma_{\mathrm{eff}}}).
\end{equation}
In the limit of strong radiation pressure cooling
$\Gamma_{\mathrm{eff}}>>\Gamma$ and cavities with $\omega_0
\tau=1$ the relation for minimal temperature eq. (\ref{eq:temp})
for a cavity illuminated with the laser power $P_{max}$ reads as
\begin{equation}
T_{\mathrm{min, rad}}=(\frac{P_0}{2mc^2\Gamma}\frac{L}{\lambda/2}
\frac{g_\mathrm{rad}^2}{\alpha_\mathrm{rad}
\beta_\mathrm{rad}})^{-1}.
\end{equation}
for radiation pressure cooling and
\begin{equation}
T_{\mathrm{min, pth}}=(n_\tau n
\frac{P_0}{2mc^2\Gamma}\frac{L}{\lambda/2}
\frac{g_\mathrm{pth}^2}{\alpha_\mathrm{pth}
\beta_\mathrm{pth}})^{-1}
\end{equation}
for photo-thermal cooling. The above derivation gives the means to
compare radiation pressure cooling with photo-thermal cooling. One
would intuitively think that the photo-thermal effect leads
ultimately to heating and that only through radiation pressure
cooling one could reach the lowest temperatures. This however
needs to be substantiated with numbers as the parameters $n$,
$n_\tau$, $\alpha$ and $\beta$ can differ in both cooling methods
by several decades. We
offer here a direct comparison.\\
Photo-thermal cooling can have a higher cooling rate than
radiation pressure cooling as long as
$T_\mathrm{pth}<T_\mathrm{rad}$, which translates into the
condition
\begin{equation}
\frac{\alpha_\mathrm{pth}\beta_\mathrm{pth}}
{g^2_\mathrm{pth}n_\tau
n}<\frac{\alpha_\mathrm{rad}\beta_\mathrm{rad}}
{g^2_\mathrm{rad}}\,.
\end{equation}
Because the absorption $\alpha$ scales as $1/g$, this can be
reformulated as
\begin{equation}
\frac{\beta_\mathrm{pth}} {g^3_\mathrm{pth}n_\tau
n}<\frac{\beta_\mathrm{rad}} {g^3_\mathrm{rad}}\,.
\end{equation}
In the case of a single experiment with competing cooling
mechanisms, we take $\beta_\mathrm{pth}=\beta_\mathrm{rad}$ and
$g_\mathrm{pth}=g_\mathrm{rad}$. Then the condition for
photo-thermal cooling to be superior over radiation pressure
cooling is $n_\tau n>1$. Typically, $n$ lies in the range of
several thousand, whereas $n_\tau$ can be designed to be as large
as $10^{10}$. To give an example for a mechanical resonator with
$f_0=100$ MHz, a delay time of $1.6\times10^{-9}$ s would be
optimal. In a cavity with $g=9$ ($R=0.8$) and cavity length $L=1$
mm, the photon storage time is only as low as $1.7\times10^{-11}$
s. If the mirror is designed in a way that $n_\tau=100$ and
$n=1000$, photo-thermal cooling is $10^5$ times more efficient
than
radiation pressure cooling.\\
More generally, if one seeks the most promising mechanism to reach
low temperatures, one would have to consider that $n$ is
proportional to the absorption $n=\xi \alpha$ with the constant
$\xi$ describing the distortion of the mirror with illumination.
Then, one needs to compare
\begin{equation}
\frac{\beta_\mathrm{pth}}
{g^2_\mathrm{pth}n_\tau\xi}\,\mathrm{with}\,\frac{\beta_\mathrm{rad}}
{g^3_\mathrm{rad}}\,.
\end{equation}
If we consider the case of $\beta_\mathrm{pth}=\beta_\mathrm{rad}$
for the sake of simplicity, one is left with a comparison of
$g^2_\mathrm{pth}n_\tau\xi$ and $g^3_\mathrm{rad}$. Should the
realization of high value of $n_\tau$ and $\xi$, which rely solely
on thermal and thermo-mechanical properties of the system, be
easier to achieve than a corresponding improvement of the optical
$g$, then photo-thermal effects would prove to be more promising
than radiation pressure to reach low temperatures and approach the
oscillator's quantum ground state. The results shown in this work
give already a guess of this trend: with a low optical finesse
cavity and with adequate thermal properties, we reached
temperatures in the range of 10 K in ref\cite{Hoeh04}, exactly as
reported more recently in very high finesse cavities for radiation
pressure\cite{Arcizet06, Schliess06}.
%
\section{conclusion}\label{sec:7}
We described a passive photo-thermal cooling mechanism. In our
experiments with gold coated micromirrors we were able to cool the
thermal vibrations of the mirror from room temperature down to the
range of 10 K. The back-action mechanism involving a photo-thermal
force time-delayed with respect to any change in mirror position
that enables this startling result was described in detail. A
theoretical account on the delay time, in our case the time of
heat conduction along the mirror, is given and shown to be in good
agreement with instantaneous and delayed response measurements. We
found that not only the lowest vibrational mode of the mirror is
cooled by optical back-action, but also higher modes as well. This
result is consistent with the theory which indicates that the
energy taken out of any vibrational mode is not transferred into
other modes but irreversibly extracted out of the vibrating
mirror. A comparison between cooling power of experiments using
radiation pressure and photo-thermal cooling is given. The
conditions for which photo-thermal cooling leads to lower
temperatures than radiation pressure cooling were specified in
detail.
\\
%
We thank T. H\"{a}nsch, T. Kippenberg and J. Kotthaus for fruitful
discussions. This work was funded by project NOMS KA 1216/2-1 of
the Deutsche Forschungsgemeinschaft (DFG). I.F. acknowledges
support of the Alexander von Humboldt Foundation.
\\
%

%

%







\end{document}